\documentclass{article}
\usepackage{spconf,amsmath,graphicx}
\usepackage{color}
\usepackage{xcolor}
\usepackage{textgreek}
\usepackage{amssymb}
\usepackage{amsthm}
\usepackage{diagbox}
\usepackage{hyperref}
\usepackage{multirow}
\usepackage{siunitx}
\usepackage{balance}
\usepackage{booktabs} 

\newcommand{\comment}[1]{}
\newcommand{\resolved}[1]{}

\title{SpectroStream: A Versatile Neural Codec for General Audio}
\name{Yunpeng Li, Kehang Han, Brian McWilliams, Zal\'an Borsos, Marco Tagliasacchi}
\address{
Google DeepMind\\
{\tt\small\{yunpeng,kehanghan,bmcw,zborsos,mtagliasacchi\}@google.com}
}
\begin{document}
%
\maketitle
\begin{abstract}

We propose SpectroStream, a full-band multi-channel neural audio codec. Successor to the well-established SoundStream, SpectroStream extends its capability beyond 24 kHz monophonic audio and enables high-quality reconstruction of 48 kHz stereo music at bit rates of 4--16 kbps. This is accomplished with a new neural architecture that leverages audio representation in the time-frequency domain, which leads to better audio quality especially at higher sample rate. The model also uses a delayed-fusion strategy to handle multi-channel audio, which is crucial in balancing per-channel acoustic quality and cross-channel phase consistency.

\end{abstract}
%

\section{Introduction}
\label{sec:intro}

A high-fidelity audio codec with a compact representation is crucial for audio compression and generation. The development of SoundStream~\cite{zeghidour2021soundstream} was a key development that unlocked the potential of autogressive audio generation with standard language-modeling objectives~\cite{borsos2023audiolm}. Since then, many new codecs have been proposed that pushed the frontier in fidelity and compactness~\cite{defossez2023encodec,kumar2023dac,defossez2024moshi,dellalibera2025focalcodec,welker2025flowdec}. However, most focus on speech~\cite{defossez2023encodec,defossez2024moshi,dellalibera2025focalcodec} or limit their attention to single-channel audio~\cite{kumar2023dac,welker2025flowdec}. This makes them either unsuitable or suboptimal for more general audio such as music, which is typically presented in a stereo format.

In this paper we propose \emph{SpectroStream}, a neural codec that can encode full-band 48 kHz general audio with joint modeling of multiple audio channels. Unlike SoundStream~\cite{zeghidour2021soundstream}, SpectroStream models the audio in the time-frequency domain with a 2D-convolutional encoder-decoder architecture. For multi-channel audio, it employs a delayed fusion and early splitting strategy in the encoder and the decoder, respectively, allowing audio channels to be processed independently in some parts of the model but jointly in other parts (see Figure~\ref{fig:generator_architecture}). This leads to more efficient encoding and decoding of multi-channel audio, as well as improved consistency across channels.

Moreover, SpectroStream uses causal convolutions with a small look-ahead in its encoder and decoder. This design choice, supported by the KWS streaming framework~\cite{rybakov2020streaming}, allows it to perform real-time streaming inference with only a small architectural latency. This can be readily achieved on a single desktop CPU, without needing specialized accelerators.

\subsection{Related Work} 
\label{sec:related-work}


SoundStream~\cite{zeghidour2021soundstream} is a convolutional encoder-decoder model based on SEANet~\cite{tagliasacchi2020} architecture. The latent representation in the bottleneck is quantized using a \emph{residual vector quantizer} (RVQ). Aside from being an efficient compressed representation for general audio, SoundStream has shown remarkable success as a representation for generating speech, general audio~\cite{borsos2023audiolm, borsos2023soundstorm} and high-fidelity polyphonic music~\cite{cideron2024musicrl, agostinelli2023musiclm}. SoundStream is trained using a combination of a waveform reconstruction loss and an adversarial loss.

Encodec~\cite{defossez2023encodec} and Descript Audio Codec (DAC)~\cite{kumar2023dac} propose methods similar to SoundStream but with modifications to the architectures and training objectives. Encodec uses a multi-scale STFT discriminator and an additional relative feature matching loss on the generator. 
DAC combines a similar discriminator with an improved RVQGAN architecture and adds a multi-scale reconstruction loss on the mel spectrogram. 

Mimi~\cite{defossez2024moshi} takes a different approach, as it was designed for generation tasks from the outset. Mimi combines the RVQ structure of SoundStream-like methods, and it additionally distills non-causal semantic information into the first level of the RVQ code. This modification is motivated by the insight that many generative models of speech combine both SoundStream-like acoustic tokens with more semantic representations to provide the model with improved linguistic capabilities.

\section{Model Architecture}
\label{sec:architecture}

\begin{figure*}[t]
\centering
\includegraphics[width=\textwidth]{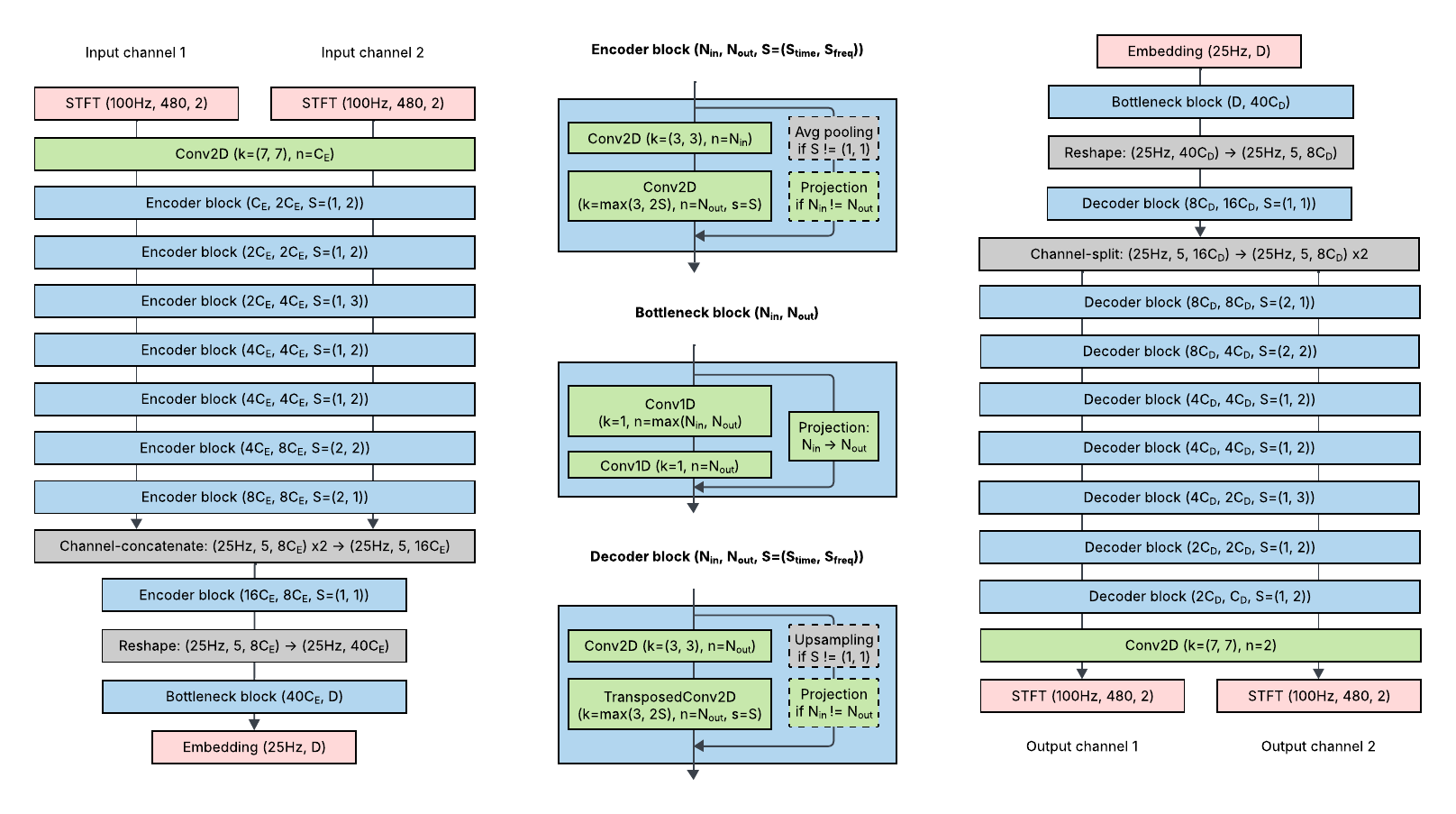}
\caption{Encoder (left) and decoder (right) architecture. The model in our experiment has base convolution depths of $C_e=32$ for the encoder and $C_d=64$ for the decoder.}
\label{fig:generator_architecture}
\end{figure*}

SpectroStream consists of an encoder, a decoder, and a quantizer. Additionally, a discriminator is used solely for adversarial learning, which we will describe in Section~\ref{sec:learning}.

The input audio is first converted into spectrograms by applying short-time Fourier transform (STFT) to each of its stereo channels. The real and imaginary components of the complex output are treated as the channel dimension of the input tensor to the encoder network. The encoder's output embedding vectors are quantized into integer tokens using residual vector quantization (RVQ)~\cite{zeghidour2021soundstream} to create a compact discrete representation of the input audio. At decoding time, the tokens are dequantized back into embeddings and inverse STFT is applied to the decoder output at the end to recreate the time-domain waveform.

The discriminator works the same way as the encoder except that it also receives the modulus of the STFT spectrogram as input, along with the real and imaginary values. Therefore, the discriminator input tensor has three channels (for each input audio channel) as opposed to just two for the encoder.

The encoder, the decoder, and the discriminator are all 2D convolutional networks that process the signal as time-frequency spectrograms. The encoder and the discriminator use strided convolutions to down-sample their inputs, while the decoder mirrors the encoder and employs transposed convolutions for up-sampling.
The encoder employs \emph{delayed fusion}: The early layers of the encoder process each input audio channel separately, with shared layer parameters but no inter-channel interaction. The intermediate tensors for the audio channels are fused just before the final few layers, which then process them jointly to produce joint embeddings for all audio channels. The delayed fusion scheme is again mirrored in the decoder (as early splitting) and emulated in the discriminator.
Finding the right fusion point is crucial for good compression and reconstruction: Fusing too early results in poor acoustic quality, as such early interactions between channels can distract the model from fidelity and realism objectives; whereas fusing too late makes the reconstructed audio lose phase consistency due to insufficient joint processing, and sound unfocused and directionless.

\begin{figure}
\centering
\includegraphics[width=0.4\textwidth]{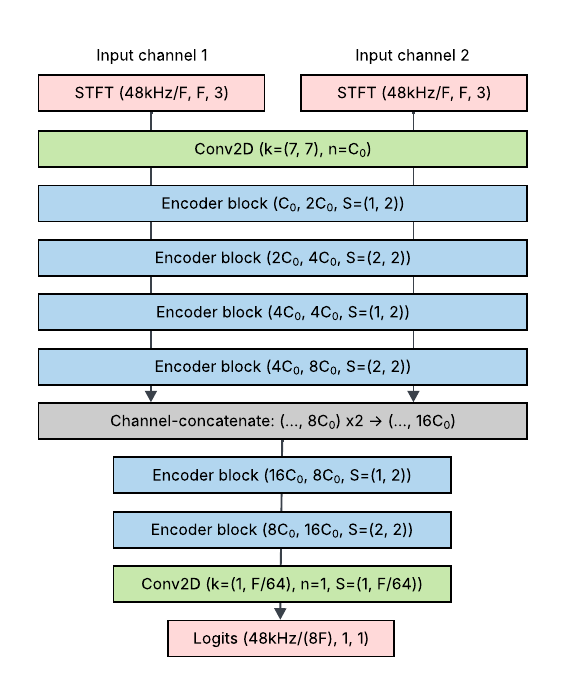}
\caption{Discriminator architecture. The number of frequency bins $F$ is half of the STFT frame length (keeping the DC and omitting the Nyquist bin), which is equal to the STFT frame step size as we use a 2x overlapping factor. We use $C_0=32$ for all discriminators in our experiments.}
\label{fig:discriminator}
\end{figure}

The full encoder and decoder architecture for stereo audio is given in Figure~\ref{fig:generator_architecture}, and the discriminator is shown in Figure~\ref{fig:discriminator}.
We use weight normalization~\cite{salimans2016} in the convolution layers of the encoder and the decoder, and layer normalization~\cite{ba2016} in the discriminator. Following the original formulation of~\cite{ba2016}, layer normalization is applied after convolution (before activation) and computed over all non-batch tensor dimensions.
The encoder and the decoder use \emph{causal} convolutions to ensure minimal latency for real-time inference. The discriminator, on the other hand, uses regular convolutions with symmetric padding, since it is not needed at inference time and so its look-ahead latency is not a consideration.
All convolutions are \emph{pre-activated} -- that is, preceded by an activation function -- with the exception of the very first layer in the network. We use ELU~\cite{clevert2016} in the encoder/decoder, and Leaky ReLU~\cite{maas2013} with a slope of $0.2$ in the discriminator. 

For 48 kHz stereo music we use an STFT window length of 960 and step size of 480, which gives us 100 STFT frames per second. We found this to be a good balance between time resolution and frequency resolution, though other choices are also possible. For the Fourier transform, we use a Hann window with no normalization or padding. To avoid the odd number of frequency bins, we omit the highest (Nyquist) frequency but keep the DC component as the latter has a lot more energy. After passing the input audio spectrogram through the strided convolutions in the encoder, we obtain $25$ embedding vectors per second.
We set the dimension of the embedding to $256$, as we found $128$ to give worse reconstruction quality and $512$ to be no better. For quantization we use $64$ residual quantizers, each with a vocabulary size of $1024$. The vocabulary, i.e., the centroids, of the quantizers are not shared. This results in a maximum bit rate of $16$ kbps. The bit rate can be set flexibly at inference time according to need by adjusting the number of quantizers used, which is a major benefit of RVQ. Furthermore, we give the decoder a one-embedding look-ahead by shifting its input relative to its output. We found this to be beneficial compared with having no look-ahead at all, while the marginal gain of even longer look-ahead is less substantial. Therefore, the small look-ahead is a good compromise between quality and latency. No look-ahead is added to the encoder, as we found it to be less effective there. This results in a total architectural latency of two embeddings, or $80$ milliseconds.
The model in our experiments has a total of $61$ million parameters at inference time: $9$M in the encoder, $36$M in the decoder, and $16$M in the quantizer.

\section{Learning}
\label{sec:learning}

\begin{figure}
\centering
\includegraphics[width=0.5\textwidth]{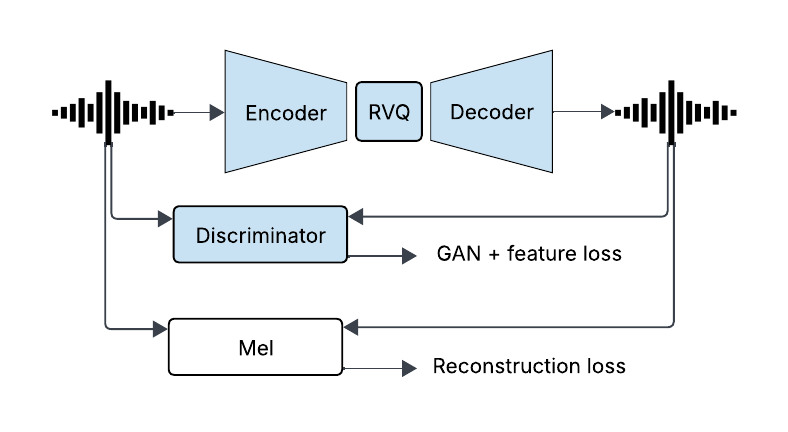}
\caption{Training setup.}
\label{fig:training_setup}
\end{figure}

The model is trained with a combination of adversarial and reconstruction losses similar to~\cite{zeghidour2021soundstream}, as illustrated in Figure~\ref{fig:training_setup}.

Unlike~\cite{zeghidour2020}, we do not use the wave-based discriminator~\cite{kumar2019melgan}. Instead, we create a multi-scale STFT-based discriminator consisting of a collection of individual discriminators, each operating at a different STFT window length. This helps the discriminator capture details at different time-frequency resolutions, analogous to the multi-period discriminator~\cite{kong2020hifigan} but on STFT spectrograms rather than raw waveform. Hence the multi-scale discriminator is
\begin{equation}
    D(x) = \frac{1}{K} \sum_{k} D_k(x),
\end{equation}
where $K$ is the number of scales. We use STFT window lengths $\left\{ 128, 256, ..., 4096 \right\}$, resulting in $K=6$ scales.
The discriminators all share the same architecture (Figure~\ref{fig:discriminator}) but do not share parameters.
Following the terminology of generative adversarial networks (GAN)~\cite{goodfellow2014gan}, we denote the encoder-quantizer-decoder modules collectively as the "generator": 
%
\begin{equation}
    G(x) = \left(\text{Dec} \circ Q^{-1} \circ Q \circ \text{Enc}\right)(x).
\end{equation}
Let $N_k$ denote the number of logits for discriminator $D_k$.
The classical GAN objective with hinge loss for the discriminator can be written as
\begin{eqnarray}
    \mathcal{L}_{D}(x) & = & \frac{1}{K} \sum_{k} \frac{1}{N_k} \sum_{n} \max \left( 0, 1 - D_{k,n}(x) \right) + \\
    \nonumber
                       &   & \frac{1}{K} \sum_{k} \frac{1}{N_k} \sum_{n} \max \left( 0, 1 + D_{k,n}(G(x)) \right),
\end{eqnarray}
while the adversarial loss for the generator is
\begin{equation}
    \mathcal{L}_{G}^{\text{adv}}(x) = \frac{1}{K} \sum_{k} \frac{1}{N_k} \sum_{n} \max \left( 0, 1 - D_{k,n}(G(x)) \right).
\end{equation}

In addition to the classical GAN loss, we include a feature loss $\mathcal{L}_{G}^{\text{feat}}$ and a reconstruction loss $\mathcal{L}_{G}^{\text{recon}}$ to ensure fidelity of the reconstructed audio to the original~\cite{tagliasacchi2020,zeghidour2021soundstream}. The feature loss encourages the reconstructed output to match the input in the feature space of the discriminator, defined by the outputs of its intermediate layers.
Let $L$ be the number of intermediate outputs, and let $M_{k, l}$ denote the size of the $l$-th intermediate output tensor of discriminator $D_k$. The feature loss is defined as
\begin{equation}
    \mathcal{L}_{G}^{\text{feat}}(x) = \mathbb{E}_{x} \left[ \frac{1}{KL} \sum_{k,l} \frac{1}{M_{k,l}} \left\| D_{k}^{(l)}(x) - D_{k}^{(l)}(G(x)) \right\|_1 \right].
\end{equation}
We use a reconstruction loss similar to~\cite{zeghidour2021soundstream} based on the mel spectral energy distance~\cite{gritsenko2020spectral}, which is given by
\begin{eqnarray}
    \mathcal{L}_{G}^{\text{rec}}(x) & = & \sum_{s} \frac{1}{B_s} \left\| S_{s}(x) - S_{s}(G(x)) \right\|_1 + \\
    \nonumber
                                      &   & \sum_{s} \frac{\alpha_{s}}{B_s} \left\| \log S_{s}(x) - \log S_{s}(G(x)) \right\|_2^2,
\end{eqnarray}
where $S_{s}(x)$ denotes the $64$-bin mel spectrogram with window length $s \in \left\{ 64, 128, ..., 2048 \right\}$ and hop length $s/4$ and $B_s$ is the size of $S_{s}(x)$. We set $\alpha_{s}$ to $\sqrt{s/2}$ following~\cite{zeghidour2021soundstream,gritsenko2020spectral}.

Finally we introduce a \emph{commitment loss} $\mathcal{L}_{G}^{\text{com}}$ to encourage the encoder output to coincide with (or "commit" to) the quantized embedding vectors, which takes the form of a squared L2 distance
\begin{equation}
    \mathcal{L}_{G}^{\text{com}}(x) = \left\| \text{Enc}(x) - \left(Q^{-1} \circ Q \circ \text{Enc}\right)(x) \right\|_2^2.
\end{equation}

The total loss of the generator is therefore a weighted sum of the losses above:
\begin{equation}
    \mathcal{L}_{G} = \lambda_\text{adv}\mathcal{L}_{G}^{\text{adv}}
                    + \lambda_\text{feat}\mathcal{L}_{G}^{\text{feat}}
                    + \lambda_\text{rec}\mathcal{L}_{G}^{\text{rec}}
                    + \lambda_\text{com}\mathcal{L}_{G}^{\text{com}}.
\end{equation}
We empirically set $\lambda_\text{feat}$ to $100$ and all other weight coefficients to $1$, and train the generator and the discriminators jointly via stochastic gradient descent using ADAM~\cite{kingma2015adam} with a learning rate of $10^{-4}$.

\subsection{Quantizer}
\label{subsec:quantizer}

Since neither quantization $Q$ nor dequantization $Q^{-1}$ is differentiable, we use a straight-through estimator by applying stop-gradients ($\text{SG}$) and substitute $Q^{-1} \cdot Q(v)$ with $v + \text{SG}(Q^{-1} \cdot Q(v) - v)$ in the backward pass. Note that gradients are not needed to train the quantizer itself, since it learns by continuously updating the exponential moving averages of its centroids.

The quantizer is trained broadly in the same methodology as~\cite{zeghidour2021soundstream}, but with the following differences in quantizer dropout.

\subsubsection{Biased quantizer dropout}
\label{subsubsec:biased-quantizer-dropout}

Recall that quantizer dropout operates during training by randomly sampling a truncation level $r \in \{1, 2, ..., R\}$ for the $R$ residual vector quantizers $(Q_1, ..., Q_R)$, and omitting all $Q_i$ for $i>r$. In~\cite{zeghidour2021soundstream}, $r$ is sampled uniformly. This allows the model to lower the number of quantizers to use at inference time, hence flexibly trading off quality for compression.

We can enhance this trade-off at the low bit rate end by biasing the sampling of $r$ towards the lower end. Specifically, we define three ranges for $r$: $\{1,...,R/4\}$, $\{R/4+1,...,R/2\}$, and $\{R/2+1,...,R\}$. We make the sampling probability for a level in the first range twice as high as in the second, and four times as high as in the third. This emulates a quasi-exponentially decreasing density.

\subsubsection{Full quantizer bypass}

As observed in~\cite{kumar2023dac}, training can be made more effective by randomly bypassing the entire quantizer altogether. Our hypothesis is that this technique provides the encoder with direct, unaltered gradients from the decoder, which are otherwise always approximated by the straight-through estimator for the quantizer. We use the same bypass rate as~\cite{kumar2023dac}, which is $0.5$. The combination of full quantizer bypass and biased dropout ensures good performance at both the high and low ends of the bit rate range.

\section{Experiments}
\label{sec:experiments}

We trained our model on a generic music dataset for 2 million steps with a batch size of 128 and training example length of 1.28 seconds. Both the generator and the discriminators are updated exactly once in each step.


We compare SpectroStream with Descript Audio Codec (DAC)~\cite{kumar2023dac}, which provides a pre-trained model.

We use the open-source implementation of ViSQOL~\cite{chinen2020visqol-v3,hines2012visqol} as our main objective quality metric. For multi-channel audio, we compute ViSQOL scores for each channel and take the average. Although ViSQOL was originally developed as a speech quality metric~\cite{hines2012visqol}, we found it to have good correlation with perceived quality for music as well.\footnote{Although an "audio mode" for general non-speech audio at a higher sample rate also exists in~\cite{chinen2020visqol-v3}, we found it to be more prone to saturation and not as informative as the so-called "speech mode".}

For this study, we selected 49 tracks from the MUSDB dataset~\cite{musdb18}, with lengths ranging from 60s to 360s. These tracks were subjected to an encode-decode process to reconstruct the audio. The perceptual similarity between the original and the reconstructed audio signals was then evaluated using ViSQOL. Both DAC and SpectroStream operate across various bit rates. We specifically examined their performances at three distinct per-channel bit rates: 2.7 kbps (low), 5.3 kbps (medium), and 8 kbps (high).

\begin{table}[h!]
\centering
\begin{tabular}{lccc}
\toprule
             & 2.7 kbps & 5.3 kbps & 8 kbps \\
\midrule
SpectroStream & \textbf{3.21}   & \textbf{3.83}    & \textbf{4.00}  \\
\midrule
DAC          & 1.47   & 2.41    & 3.33  \\
\bottomrule
\end{tabular}
\caption{\label{tab:visqol_eval} ViSQOL scores for SpectroStream and DAC at various per-channel bit rates. Higher ViSQOL scores indicate superior perceptual audio quality.}
\end{table}

As presented in Table \ref{tab:visqol_eval}, SpectroStream consistently demonstrates superior performance compared to DAC across all evaluated bit rates. Notably, at the low $2.7$ kbps, SpectroStream achieves a ViSQOL gain exceeding $1.7$ points.

We conducted a subjective listening test to assess perceptual quality. This involved 60 trained raters evaluating 20 distinct audio clips, also sourced from the MUSDB dataset~\cite{musdb18}, with each clip approximately 10 seconds in length. The evaluation employed an A/B comparison methodology, where raters judged the perceived quality of SpectroStream against DAC, with the original reference audio serving as a benchmark. As shown in Table \ref{tab:human_eval}, the results of the subjective evaluation align with the objective metrics. While both models do well on the highest bit rate, where the difference is often too subtle to be obvious, the preference for SpectroStream is much more pronounced at low bit rates, where listeners preferred SpectroStream over DAC 76.3\% of the time.

\begin{table}[h!]
\centering


\begin{tabular}{lccc}
\toprule
             & 2.7 kbps & 5.3 kbps & 8 kbps \\
\midrule
SpectroStream & \textbf{76.3\%}   & \textbf{55.0\%}    & \textbf{50.8\%}  \\
\midrule
DAC          & 23.7\%   & 45.0\%    & 49.2\%  \\
\bottomrule
\end{tabular}
\caption{\label{tab:human_eval} Subjective evaluation results: SpectroStream's win rate over DAC at various per-channel bit rates. Values represent the percentage of times SpectroStream was preferred over DAC. Higher win rates indicate superior perceptual audio quality for SpectroStream.}
\end{table}

\section{Conclusion}
\label{sec:conclusion}

We have presented SpectroStream, a full-band, multi-channel neural audio codec designed for high-quality 48 kHz stereo music. A key innovation of our model lies in its 2D convolutional architecture, which operates directly on the time-frequency representation. This, combined with a multi-scale STFT-based discriminator, leads to significant improvements in reconstruction quality. For multi-channel audio, we introduced a delayed-fusion and early-splitting strategy that effectively balances per-channel fidelity with cross-channel phase coherence, a critical aspect for a natural stereo listening experience.

Experimental results show that our model compares favorably against state-of-the-art models with similar capabilities, both objectively and subjectively. The advantage is particularly pronounced at lower bit rates.
The high-fidelity yet compact representation produced by SpectroStream provides a stronger foundation for language modeling on audio, and paves the way for more sophisticated audio generation applications.

\balance
\bibliographystyle{IEEEtran}
{
\small  
\bibliography{references}

\begin{thebibliography}{10}
\providecommand{\url}[1]{#1}
\csname url@samestyle\endcsname
\providecommand{\newblock}{\relax}
\providecommand{\bibinfo}[2]{#2}
\providecommand{\BIBentrySTDinterwordspacing}{\spaceskip=0pt\relax}
\providecommand{\BIBentryALTinterwordstretchfactor}{4}
\providecommand{\BIBentryALTinterwordspacing}{\spaceskip=\fontdimen2\font plus
\BIBentryALTinterwordstretchfactor\fontdimen3\font minus
  \fontdimen4\font\relax}
\providecommand{\BIBforeignlanguage}[2]{{%
\expandafter\ifx\csname l@#1\endcsname\relax
\typeout{** WARNING: IEEEtran.bst: No hyphenation pattern has been}%
\typeout{** loaded for the language `#1'. Using the pattern for}%
\typeout{** the default language instead.}%
\else
\language=\csname l@#1\endcsname
\fi
#2}}
\providecommand{\BIBdecl}{\relax}
\BIBdecl

\bibitem{zeghidour2021soundstream}
\BIBentryALTinterwordspacing
N.~Zeghidour, A.~Luebs, A.~Omran, J.~Skoglund, and M.~Tagliasacchi,
  ``Soundstream: An end-to-end neural audio codec,'' \emph{IEEE/ACM
  Transactions on Audio, Speech, and Language Processing}, vol.~30, pp.
  495--507, 2021. [Online]. Available:
  \url{https://ieeexplore.ieee.org/stamp/stamp.jsp?arnumber=9625818}
\BIBentrySTDinterwordspacing

\bibitem{borsos2023audiolm}
Z.~Borsos, R.~Marinier, D.~Vincent, E.~Kharitonov, O.~Pietquin, M.~Sharifi,
  D.~Roblek, O.~Teboul, D.~Grangier, M.~Tagliasacchi, and N.~Zeghidour,
  ``Audiolm: A language modeling approach to audio generation,'' \emph{IEEE
  Transactions on Audio, Speech, and Language Processing}, vol.~31, pp.
  2523--2533, 2023.

\bibitem{defossez2023encodec}
A.~D{\'e}fossez, J.~Copet, G.~Synnaeve, and Y.~Adi, ``High fidelity neural
  audio compression,'' \emph{Transactions on Machine Learning Research}, 2023.

\bibitem{kumar2023dac}
R.~Kumar, P.~Seetharaman, A.~Luebs, I.~Kumar, and K.~Kumar, ``High-fidelity
  audio compression with improved {RVQGAN},'' in \emph{Neural Information
  Processing Systems}, 2023.

\bibitem{defossez2024moshi}
A.~Défossez, L.~Mazaré, M.~Orsini, A.~Royer, P.~Pérez, H.~Jégou, E.~Grave,
  and N.~Zeghidour, ``Moshi: a speech-text foundation model for real-time
  dialogue,'' 2024.

\bibitem{dellalibera2025focalcodec}
L.~D. Libera, F.~Paissan, C.~Subakan, and M.~Ravanelli, ``Focalcodec:
  Low-bitrate speech coding via focal modulation networks,'' 2025.

\bibitem{welker2025flowdec}
\BIBentryALTinterwordspacing
S.~Welker, M.~Le, R.~T.~Q. Chen, W.-N. Hsu, T.~Gerkmann, A.~Richard, and Y.-C.
  Wu, ``{FlowDec}: A flow-based full-band general audio codec with high
  perceptual quality,'' in \emph{The Thirteenth International Conference on
  Learning Representations}, 2025. [Online]. Available:
  \url{https://openreview.net/forum?id=uxDFlPGRLX}
\BIBentrySTDinterwordspacing

\bibitem{rybakov2020streaming}
\BIBentryALTinterwordspacing
O.~Rybakov, N.~Kononenko, N.~Subrahmanya, M.~Visontai, and S.~Laurenzo,
  ``Streaming keyword spotting on mobile devices,'' \emph{CoRR}, 2020.
  [Online]. Available: \url{https://arxiv.org/abs/2005.06720}
\BIBentrySTDinterwordspacing

\bibitem{tagliasacchi2020}
M.~Tagliasacchi, Y.~Li, K.~Misiunas, and D.~Roblek, ``{SEANet}: A multi-modal
  speech enhancement network,'' in \emph{INTERSPEECH}, 2020.

\bibitem{borsos2023soundstorm}
Z.~Borsos, M.~Sharifi, D.~Vincent, E.~Kharitonov, N.~Zeghidour, and
  M.~Tagliasacchi, ``Soundstorm: Efficient parallel audio generation,''
  \emph{arXiv preprint arXiv:2305.09636}, 2023.

\bibitem{cideron2024musicrl}
G.~Cideron, S.~Girgin, M.~Verzetti, D.~Vincent, M.~Kastelic, Z.~Borsos,
  B.~McWilliams, V.~Ungureanu, O.~Bachem, O.~Pietquin \emph{et~al.}, ``Musicrl:
  Aligning music generation to human preferences,'' \emph{arXiv preprint
  arXiv:2402.04229}, 2024.

\bibitem{agostinelli2023musiclm}
A.~Agostinelli, T.~I. Denk, Z.~Borsos, J.~Engel, M.~Verzetti, A.~Caillon,
  Q.~Huang, A.~Jansen, A.~Roberts, M.~Tagliasacchi \emph{et~al.}, ``Musiclm:
  Generating music from text,'' \emph{arXiv preprint arXiv:2301.11325}, 2023.

\bibitem{salimans2016}
T.~Salimans and D.~P. Kingma, ``Weight normalization: A simple
  reparameterization to accelerate training of deep neural networks,'' in
  \emph{Advances in Neural Information Processing Systems}, 2016, pp. 901--909.

\bibitem{ba2016}
J.~L. Ba, J.~R. Kiros, and G.~E. Hinton, ``Layer normalization,'' \emph{arXiv
  preprint arXiv:1607.06450}, 2016.

\bibitem{clevert2016}
D.-A. Clevert, T.~Unterthiner, and S.~Hochreiter, ``Fast and accurate deep
  network learning by exponential linear units ({ELUs}),'' in
  \emph{International Conference on Learning Representations}, 2016.

\bibitem{maas2013}
A.~L. Maas, A.~Y. Hannun, and A.~Y. Ng, ``Rectifier nonlinearities improve
  neural network acoustic models,'' in \emph{ICML Workshop on Deep Learning for
  Audio, Speech and Language Processing}, 2013.

\bibitem{zeghidour2020}
\BIBentryALTinterwordspacing
N.~Zeghidour and D.~Grangier, ``Wavesplit: End-to-end speech separation by
  speaker clustering,'' \emph{CoRR}, vol. abs/2002.08933, 2020. [Online].
  Available: \url{https://arxiv.org/abs/2002.08933}
\BIBentrySTDinterwordspacing

\bibitem{kumar2019melgan}
K.~Kumar, R.~Kumar, T.~de~Boissiere, L.~Gestin, W.~Z. Teoh, J.~Sotelo,
  A.~de~Brebisson, Y.~Bengio, and A.~Courville, ``{MelGAN}: Generative
  adversarial networks for conditional waveform synthesis,'' in \emph{Advances
  in Neural Information Processing Systems}, 2019.

\bibitem{kong2020hifigan}
\BIBentryALTinterwordspacing
J.~Kong, J.~Kim, and J.~Bae, ``Hifi-gan: Generative adversarial networks for
  efficient and high fidelity speech synthesis,'' in \emph{Advances in Neural
  Information Processing Systems}, H.~Larochelle, M.~Ranzato, R.~Hadsell,
  M.~Balcan, and H.~Lin, Eds., vol.~33.\hskip 1em plus 0.5em minus 0.4em\relax
  Curran Associates, Inc., 2020, pp. 17\,022--17\,033. [Online]. Available:
  \url{https://proceedings.neurips.cc/paper_files/paper/2020/file/c5d736809766d46260d816d8dbc9eb44-Paper.pdf}
\BIBentrySTDinterwordspacing

\bibitem{goodfellow2014gan}
I.~Goodfellow, J.~Pouget-Abadie, M.~Mirza, B.~Xu, D.~Warde-Farley, S.~Ozair,
  A.~Courville, and Y.~Bengio, ``Generative adversarial nets,'' in \emph{Neural
  Information Processing Systems (NIPS)}, 2014, pp. 2672--2680.

\bibitem{gritsenko2020spectral}
A.~Gritsenko, T.~Salimans, R.~van~den Berg, J.~Snoek, and N.~Kalchbrenner, ``A
  spectral energy distance for parallel speech synthesis,'' in \emph{Advances
  in Neural Information Processing Systems}, H.~Larochelle, M.~Ranzato,
  R.~Hadsell, M.~Balcan, and H.~Lin, Eds., vol.~33.\hskip 1em plus 0.5em minus
  0.4em\relax Curran Associates, Inc., 2020, pp. 13\,062--13\,072.

\bibitem{kingma2015adam}
D.~P. Kingma and J.~Ba, ``Adam: A method for stochastic optimization.'' in
  \emph{International Conference on Learning Representations}, Y.~Bengio and
  Y.~LeCun, Eds., 2015.

\bibitem{chinen2020visqol-v3}
\BIBentryALTinterwordspacing
M.~Chinen, F.~S.~C. Lim, J.~Skoglund, N.~Gureev, F.~O'Gorman, and A.~Hines,
  ``Visqol v3: An open source production ready objective speech and audio
  metric,'' 2020. [Online]. Available: \url{https://arxiv.org/abs/2004.09584}
\BIBentrySTDinterwordspacing

\bibitem{hines2012visqol}
A.~Hines, J.~Skoglund, A.~Kokaram, and N.~Harte, ``Visqol: The virtual speech
  quality objective listener,'' in \emph{IWAENC 2012; International Workshop on
  Acoustic Signal Enhancement}, 2012.

\bibitem{musdb18}
\BIBentryALTinterwordspacing
Z.~Rafii, A.~Liutkus, F.-R. St{\"o}ter, S.~I. Mimilakis, and R.~Bittner, ``The
  {MUSDB18} corpus for music separation,'' Dec. 2017. [Online]. Available:
  \url{https://doi.org/10.5281/zenodo.1117372}
\BIBentrySTDinterwordspacing

\end{thebibliography}
}

\end{document}